\documentclass[conference]{IEEEtran}

\usepackage{cite}
\usepackage{amsmath,amssymb}
\usepackage{graphicx}
\usepackage{booktabs}
\usepackage{multirow}
\usepackage{xcolor}
\usepackage{rotating}
\usepackage{booktabs}
\usepackage{array}
\usepackage{dblfloatfix}
\usepackage{placeins}

\title{When Models Ignore Definitions: Measuring Semantic Override Hallucinations in LLM Reasoning}

\author{
\IEEEauthorblockN{
Yogeswar Reddy Thota\IEEEauthorrefmark{1},
Setareh Rafatirad\IEEEauthorrefmark{2},
Houman Homayoun\IEEEauthorrefmark{2},
Tooraj Nikoubin\IEEEauthorrefmark{1}
}

\IEEEauthorblockA{\IEEEauthorrefmark{1}
Dept. of Electrical and Computer Engineering,
University of Texas at Dallas, Richardson, TX, USA\\
\{yogeswarreddy.thota, tooraj.nikoubin\}@utdallas.edu
}

\IEEEauthorblockA{\IEEEauthorrefmark{2}
Dept. of Electrical and Computer Engineering,
University of California Davis, CA, USA\\
\{srafatirad, hhomayoun\}@ucdavis.edu
}
}

\begin{document}
\maketitle

\begin{abstract}
Large language models (LLMs) demonstrate strong performance on standard digital logic and Boolean reasoning tasks, yet their reliability under \emph{locally redefined semantics} remains poorly understood. In many formal settings, such as circuit specifications, examinations, and hardware documentation, operators and components are explicitly redefined within narrow scope. Correct reasoning in these contexts requires models to temporarily suppress globally learned conventions in favor of prompt-local definitions. In this work, we study a systematic failure mode we term \emph{semantic override}, in which an LLM reverts to its pretrained default interpretation of operators or gate behavior despite explicit redefinition in the prompt. We also identify a related class of errors, \emph{assumption injection}, where models commit to unstated hardware semantics when critical details are underspecified, rather than requesting clarification. We introduce a compact micro-benchmark of 30 logic and digital-circuit reasoning tasks designed as verifier-style traps, spanning Boolean algebra, operator overloading, redefined gates, and circuit-level semantics. Evaluating three frontier LLMs, we observe persistent noncompliance with local specifications, confident but incompatible assumptions, and dropped constraints even in elementary settings. Our findings highlight a gap between surface-level correctness and specification-faithful reasoning, motivating evaluation protocols that explicitly test local unlearning and semantic compliance in formal domains.
\end{abstract}


\section{Introduction}

Large language models (LLMs) have recently shown impressive capabilities across mathematics, programming, and formal reasoning tasks, including Boolean algebra and digital logic \cite{wei2022cot,ouyang2022instructgpt}. When evaluated on standard textbook-style problems, these models often achieve near-perfect accuracy. Recent large-scale evaluations, however, show that high accuracy on canonical tasks does not reliably translate to robust logical reasoning, particularly under structured or adversarial settings~\cite{10870148,10.24963/ijcai.2024/693}.
 However, correctness on canonical problems does not necessarily imply reliability in settings where the semantics of operators, gates, or signals are \emph{locally redefined} or partially specified.

In digital logic and circuit design, local specification overrides are common. Examination problems may redefine operators for a specific question, hardware documentation may introduce nonstandard gate behavior, and circuit-level reasoning often depends on unstated but critical assumptions about signal polarity, reset behavior, or electrical implementation (e.g., wired logic versus tri-state buses). Human solvers are expected to prioritize the prompt-local specification over globally learned conventions \cite{wallace2024instructionhierarchy,zhou2023ifeval}.

In this paper, we argue that many LLM errors in formal reasoning are not best characterized as random hallucinations~\cite{ji2023hallucinationsurvey}, but instead arise from a systematic failure to \emph{locally unlearn} strong pretrained priors. We term this phenomenon \emph{semantic override}. Under semantic override, a model disregards or underweights an explicit prompt level redefinition such as a gate labeled “NAND” behaving as AND or an operator redefined to mean modulo-2 addition and instead reverts to its globally learned interpretation.

We further identify a complementary failure mode, which we term \emph{assumption injection}. In underspecified problems such as circuit questions lacking explicit reset polarity, clock edge, or bus implementation models frequently commit to a single interpretation without acknowledging ambiguity or requesting clarification \cite{lin2022truthfulqa}. While such assumptions may be reasonable in isolation, confident commitment without justification leads to incompatible answers across models and violates the verifier-like behavior expected in safety-critical or specification-driven contexts.

To study these behaviors in a controlled and reproducible manner, we introduce a focused micro-benchmark of 30 digital logic and circuit reasoning tasks. Each task is constructed as a verifier-style trap, where the correct response may require obeying a local semantic override, rejecting a contradictory premise, or explicitly flagging missing information. The benchmark spans Boolean algebra, operator overloading, redefined logic gates, sequential elements, bus semantics, and Karnaugh map optimization with don’t-care conditions.

We used EduArena a testbed of Turing company for various versions of LLM models to evaluate models in one place. Our evaluation across three frontier LLMs reveals that semantic override and assumption injection persist even in elementary settings, often accompanied by high confidence and fluent explanations. These findings suggest a gap between apparent reasoning competence and true specification compliance. By isolating and quantifying this gap, our work contributes a concrete evaluation lens for studying reliability, instruction adherence, and local unlearning in LLM's.

The remainder of this paper is organized as follows. Section~\ref{sec:related} reviews related work on hallucination, instruction following, and formal reasoning reliability. Section~\ref{sec:method} describes the benchmark design, verifier-style scoring protocol, and error taxonomy. Section~\ref{sec:results} presents experimental results and analysis, and Section~\ref{sec:conclusion} concludes with key takeaways and directions for future work.

\section{Related Work}
\label{sec:related}

The reliability of LLMs under ambiguous, underspecified, or conflicting instructions has received increasing attention in recent years. Prior work has examined hallucination, instruction following, and formal reasoning accuracy from multiple perspectives. However, the specific failure mode we study \emph{semantic override under prompt-local redefinitions} has not been explicitly isolated or systematically evaluated. 

\subsection{Hallucination and Faithfulness}
Hallucination in natural language generation broadly refers to outputs that are fluent but ungrounded, incorrect, or inconsistent with provided inputs. Ji et al.~\cite{ji2023hallucinationsurvey} provide a comprehensive taxonomy of hallucination phenomena across tasks such as summarization, question answering, and dialogue. While much of this literature emphasizes factual inaccuracies, our work highlights a distinct class of failures in formal reasoning settings, where outputs are logically coherent yet violate prompt-local semantic constraints.

SelfCheckGPT~\cite{manakul2023selfcheckgpt} and related approaches detect hallucination via response instability across generations. However, such methods are less effective at identifying \emph{stable but specification-incompatible} behavior, which is central to the semantic override errors we observe.

\subsection{Instruction Following and Rule Compliance}
Instruction-following behavior has been extensively studied in the context of alignment and reinforcement learning from human feedback. InstructGPT~\cite{ouyang2022instructgpt} demonstrates improvements in adherence to user instructions, but largely assumes that instructions are globally consistent with pretrained knowledge.

More recent evaluations explicitly probe instruction compliance under verifiable constraints. IFEval~\cite{zhou2023ifeval} shows that even state-of-the-art LLMs frequently violate explicit, machine-checkable instructions. Similarly, Liu et al.~\cite{liu2023evaluating} analyze rule-following failures in structured tasks, revealing that models often produce confident outputs that disregard stated rules. Our work extends these findings by focusing on \emph{semantic conflicts} between pretrained operator meanings and prompt-local redefinitions within formal logic domains.

\subsection{Truthfulness, False Premises, and Assumption Injection}
TruthfulQA~\cite{lin2022truthfulqa} demonstrates that LLMs tend to repeat plausible but incorrect statements rather than challenge false premises. While this benchmark focuses on natural language misconceptions, we observe an analogous phenomenon in digital logic and circuit reasoning: models frequently accept invalid axioms or inject unstated assumptions instead of flagging underspecification.

Prior work on clarification in question answering~\cite{rao2018learning} establishes that requesting additional information under uncertainty is a desirable and measurable behavior. In contrast, our experiments show that LLMs often commit to specific hardware or semantic interpretations without justification, a failure mode we term \emph{assumption injection}.

\subsection{Logical Reasoning Benchmarks}
Several recent benchmarks have systematically evaluated the logical reasoning capabilities of large language models beyond surface-level accuracy. Xu et al.~\cite{10870148} present a study analyzing deductive, inductive, and abductive reasoning, concluding that even state-of-the-art models exhibit significant logical inconsistencies under controlled evaluation. 

LogicBench~\cite{parmar2024logicbenchsystematicevaluationlogical} and Multi-LogiEval~\cite{patel2024multilogievalevaluatingmultisteplogical} further demonstrate that LLM performance degrades sharply as logical structure and multi-step dependencies increase. Recent work has also examined LLM performance in domain-specific digital logic reasoning settings. Recent work has examined LLM performance in domain-specific digital logic reasoning. Thota et al.~\cite{thota2026humanai} evaluate GPT, Gemini, and Claude on undergraduate circuit and timing-diagram problems, identifying a pronounced gap between perceived helpfulness and formal correctness. None of the models matched official solutions on the most sequentially complex tasks, despite generating confident and well-structured explanations. The study attributes these failures to systematic errors in state evolution and timing analysis, particularly in non-standard counters and finite-state machines.
While these benchmarks and domain-specific studies~\cite{thota2026humanai} evaluate reasoning accuracy under fixed operator semantics, this work complements them by probing robustness under \emph{prompt-local semantic redefinitions}, a setting common in formal specifications and digital logic reasoning.

\subsection{Positioning of This Work}
In contrast to prior work, our contribution is threefold:
\begin{itemize}
  \item We isolate \emph{semantic override} as a distinct failure mode arising from conflicts between pretrained knowledge and prompt-local definitions.
  \item We treat clarification and refusal as \emph{correct} behavior under underspecification, adopting a verifier-oriented evaluation lens.
  \item We introduce a compact, domain-specific benchmark in digital logic and circuit reasoning that exposes specification-level failures not captured by existing logical reasoning benchmarks~\cite{10870148,parmar2024logicbenchsystematicevaluationlogical}.

\end{itemize}

Together, these contributions position our work as a targeted study of specification-faithful reasoning and local unlearning in modern LLMs.

\section{Methodology: Evaluating Semantic Override Under Local Redefinitions}
\label{sec:method}

\subsection{Problem Framing}
LLMs demonstrate strong competence on canonical digital-logic and Boolean-algebra tasks. However, this study targets a narrower failure mode: \emph{semantic override}, in which a model reverts to globally learned meanings (e.g., standard gate behavior or operator semantics) despite an explicit prompt-local redefinition. This behavior reflects instruction-priority conflicts and robustness failures under conflicting specifications \cite{wallace2024instructionhierarchy}, and manifests as confident but specification-incompatible reasoning rather than random hallucination \cite{ji2023hallucinationsurvey}.

We operationalize this as a \textbf{Verifier vs. Solver} setting. The \emph{Solver} is asked to produce an answer directly. The \emph{Verifier} objective is to (i) respect locally scoped definitions, (ii) detect underspecification or contradiction, and (iii) request clarification when required rather than guessing.

\subsection{Trap-Set Construction}
\label{subsec:trapset}
We construct a curated set of $N=30$ prompts in digital logic, Boolean algebra, and circuit semantics. Each prompt is designed to be short, mechanically checkable, and to isolate a specific failure mode. Prompts are grouped into five trap families:

\begin{itemize}
  \item \textbf{Definition Override (DO):} Redefines an operator or a gate (e.g., ``$\otimes$ means OR'', ``NAND is defined as AND'') and tests whether the model follows the local definition or reverts to the standard meaning.
  \item \textbf{Hardware Semantics Ambiguity (HSA):} Uses circuit terms (e.g., wired-OR, open-drain, floating inputs) where correct resolution depends on physical assumptions. The correct behavior is either a conditional answer with explicit assumptions or a request for missing implementation details.
  \item \textbf{Underspecification (US):} Provides insufficient information to determine a unique result (e.g., partial truth table constraints), where the gold response is to state non-identifiability or enumerate valid possibilities.
  \item \textbf{Contradiction Handling (CH):} Introduces conflicting axioms or self-referential constraints; the gold response is to flag inconsistency or provide satisfiable conditions rather than forcing an unconditional solution.
  \item \textbf{Task Misread / Goal Drift (TM):} Prompts where models often shift the task (e.g., simplifying an expression vs. judging a statement as ``true''), scored as incorrect when the produced output does not address the requested form.
\end{itemize}

\subsection{Models and Prompting Protocol}
We evaluate three frontier LLMs (denoted Model-A, Model-B, Model-C) under identical prompts, with no tool use and no external resources. Each prompt is issued in a single-turn format. To reduce variance, we keep temperature fixed (or the default deterministic setting when exposed) and record the first complete response.

\subsection{Gold Labels and Verifier-Correct Criteria}
\label{subsec:gold}
Each item is annotated with a gold label specifying whether the prompt admits a unique solution under the provided local semantics (\textbf{Solve}) or is inherently underspecified, ambiguous, or contradictory (\textbf{Flag}).

For \textbf{Solve} items, verifier-correct responses must produce the correct value or expression under the prompt-local definitions. For \textbf{Flag} items, verifier-correct responses must explicitly identify missing information, ambiguity, or inconsistency. Responses that commit to a single interpretation without justification are graded incorrect.

\subsection{Error Taxonomy (What Counts as ``Went Wrong'')}
\label{subsec:errors}
We score \emph{all} incorrect behaviors, not only hallucinations. Each incorrect response is assigned one primary error type:

\begin{itemize}
  \item \textbf{E1: Semantic Override}   ignores an explicit local redefinition and uses the canonical meaning.
  \item \textbf{E2: Assumption Injection / False Commitment}   commits to a specific interpretation or physical implementation without stating assumptions, when multiple are plausible.
  \item \textbf{E3: Constraint Dropping}   performs algebraic manipulation but fails to enforce constraints when substituting back (e.g., self-referential equations).
  \item \textbf{E4: Task Misread}   answers a different question than asked (e.g., truth of a statement vs simplified form).
  \item \textbf{E5: Non-Sequitur Generalization}   introduces non-requested frameworks (e.g., ``trivial algebra where $1=0$'') without justification from the prompt.
\end{itemize}

\subsection{Scoring}
Each response is labeled as \emph{verifier-correct} or \emph{verifier-incorrect} according to the criteria in Section~\ref{subsec:gold}. All incorrect responses are further assigned a single primary error type from the taxonomy in Section~\ref{subsec:errors} (E1–E5). We report overall verifier-correct accuracy, accuracy by trap family, and error-type distributions, with particular emphasis on semantic override (E1) as an indicator of failure to suppress pretrained operator semantics under local redefinition.

\subsection{Reproducibility Package}
For transparency, we publish the full prompt set, gold labels (Solve vs Flag), gold solutions/conditions, and per-model outputs. Each prompt is tagged with its trap family and primary expected failure mode to facilitate follow-up studies and extension to additional models.
\begin{table}[t]
\centering
\caption{Overall verifier-correct accuracy on the 30-question benchmark.}
\label{tab:overall_accuracy}
\begin{tabular}{l c}
\toprule
\textbf{Model} & \textbf{Accuracy (\%)} \\
\midrule
Model-A & 90.0 \\
Model-B & 90.0 \\
Model-C & 80.0 \\

\bottomrule
\end{tabular}
\end{table}
\section{Results}
\label{sec:results}

We evaluate three frontier large language models Claude (Haiku 4.5), ChatGPT-5.2 Pro, and Gemini 3 Pro hereafter referred to as Model-A, Model-B, and Model-C, on the proposed 30-item verifier-style benchmark. Each response is graded according to the criteria in Section~\ref{sec:method}, yielding binary correctness scores and a primary error label when incorrect.

\subsection{Overall Accuracy}
Table~\ref{tab:overall_accuracy} reports overall verifier-correct accuracy across all 30 items. A response is considered correct only if it satisfies the verifier criteria defined in Section~\ref{subsec:gold}.

Despite strong performance on standard logic tasks, none of the evaluated models achieves perfect accuracy under verifier-style scoring. Errors are not uniformly distributed, but instead cluster around specific trap families.

\subsection{Accuracy by Trap Family}
\begin{figure}[t]
  \centering
  \includegraphics[width=\linewidth]{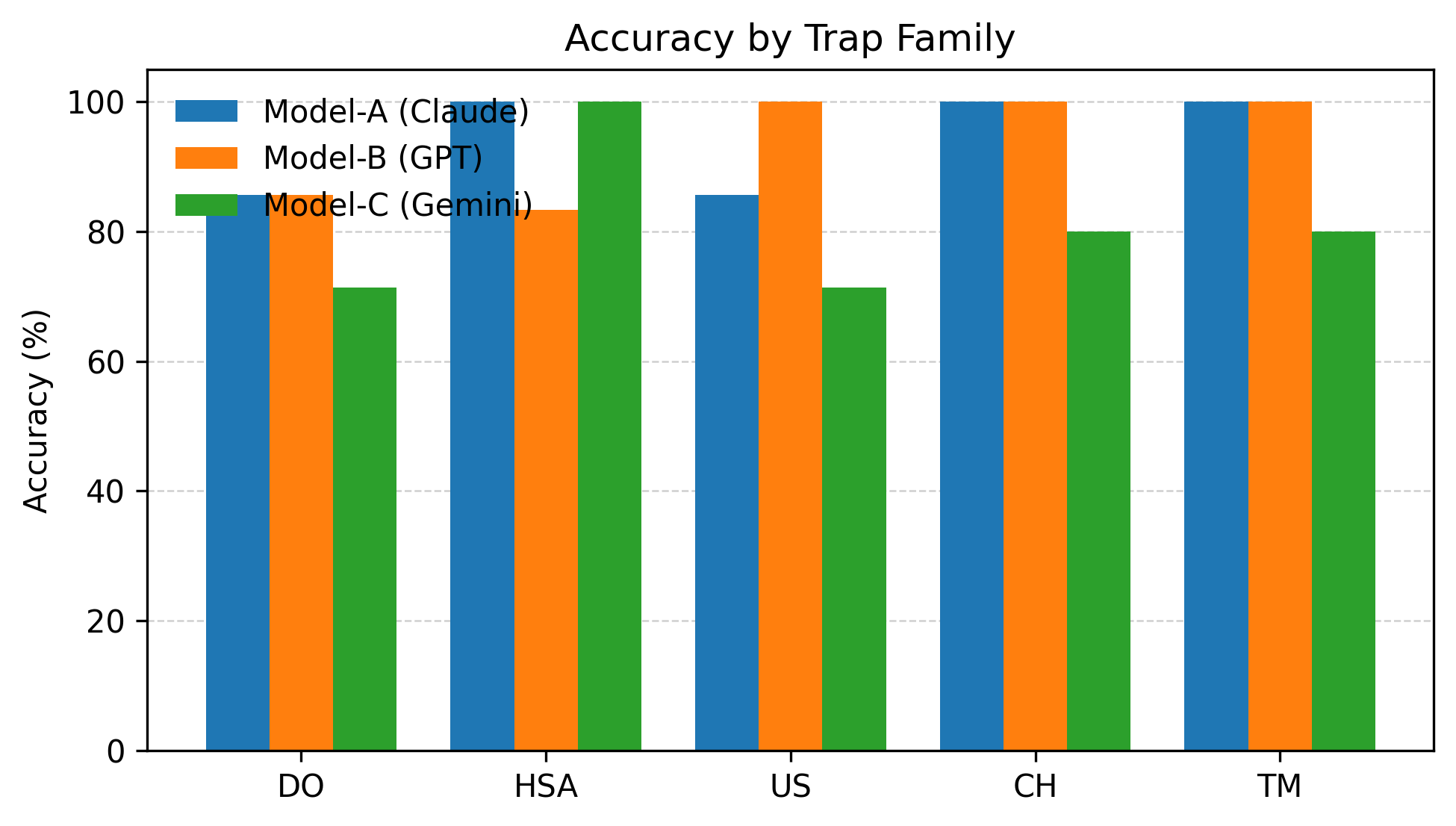}
  \caption{Accuracy by trap family across three models.}
  \label{fig:trap_accuracy}
\end{figure}

Figure~\ref{fig:trap_accuracy} summarizes accuracy broken down by trap family (Definition Override, Hardware Semantics Ambiguity, Underspecification, Contradiction Handling, and Task Misread).

\begin{table}[t]
\centering
\caption{Accuracy (\%) by trap family.}
\label{tab:trap_accuracy}
\begin{tabular}{@{}lccc@{}}
\toprule
\textbf{Trap Family} & \textbf{Model-A} & \textbf{Model-B} & \textbf{Model-C} \\
\midrule
Definition Override (DO) & 85.7 & 85.7 & 71.4 \\
Hardware Semantics Ambiguity (HSA) & 100.0 & 83.3 & 100.0 \\
Underspecification (US) & 85.7 & 100.0 & 71.4 \\
Contradiction Handling (CH) & 100.0 & 100.0 & 80.0 \\
Task Misread (TM) & 100.0 & 100.0 & 80.0 \\

\bottomrule
\end{tabular}
\end{table}

Across all three models, Definition Override (DO) traps exhibit the lowest accuracy, indicating persistent difficulty in suppressing globally learned operator semantics in favor of local redefinitions.

\begin{table}[t]
\centering
\caption{Primary error types (percentage of incorrect responses).}
\label{tab:error_distribution}
\begin{tabular}{@{}lccc@{}}
\toprule
\textbf{Error Type} & \textbf{Model-A} & \textbf{Model-B} & \textbf{Model-C} \\
\midrule
E1: Semantic Override & 33.3 & 33.3 & 33.3 \\
E2: Assumption Injection & 0.0 & 0.0 & 33.3 \\
E3: Constraint Dropping & 0.0 & 33.3 & 16.7 \\
E4: Task Misread & 0.0 & 0.0 & 16.7 \\
E5: Non-Sequitur Generalization & 0.0 & 0.0 & 0.0 \\

\bottomrule
\end{tabular}
\end{table}

\subsection{Error-Type Distribution}
\begin{figure}[t]
  \centering
  \includegraphics[width=\linewidth]{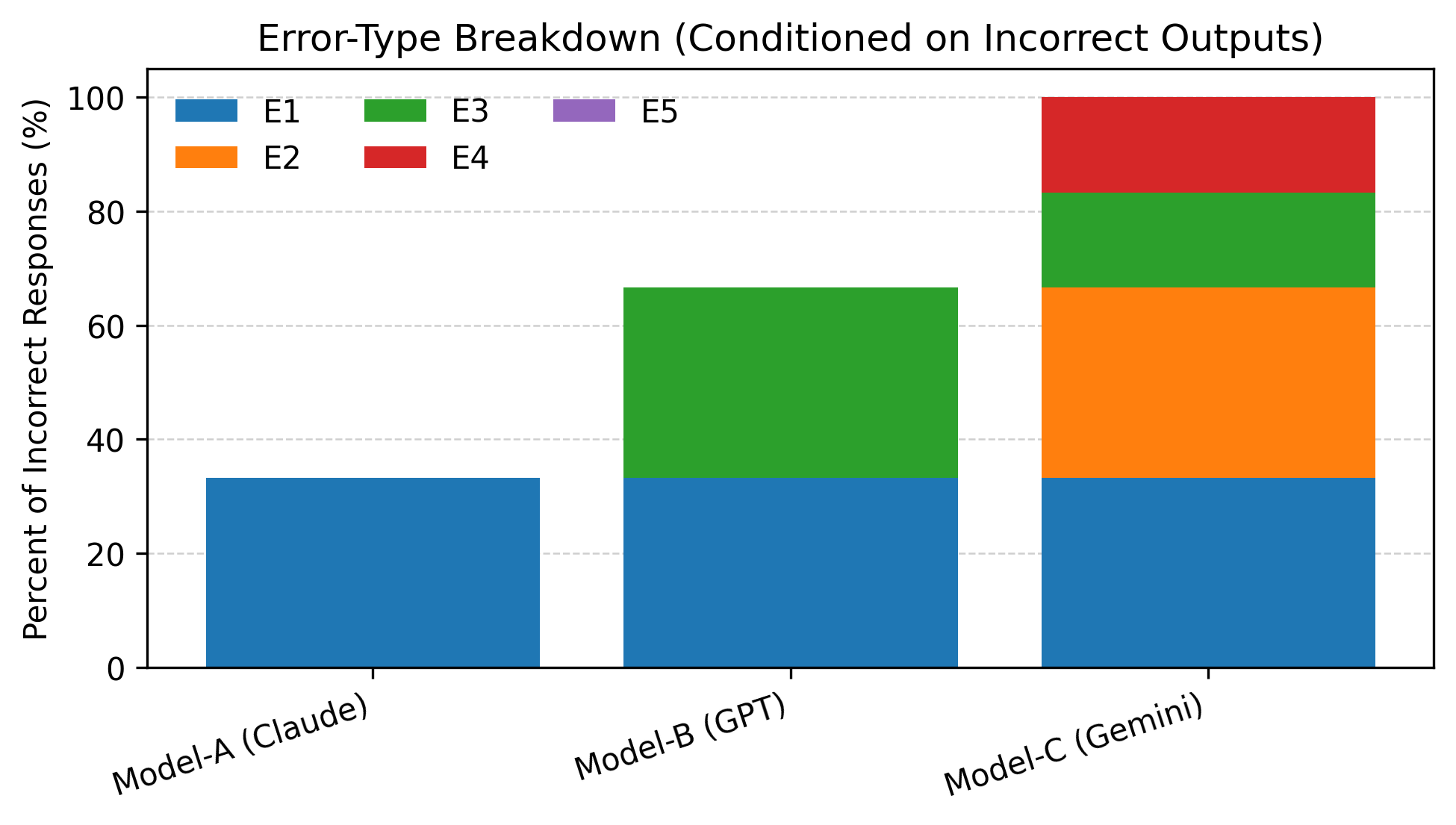}
  \caption{Distribution of verifier-incorrect responses across error types (E1--E5) for each model.}
  \label{fig:error_distribution}
\end{figure}

To better understand the nature of model failures, we analyze the distribution of primary error types (E1--E5; see Section~\ref{subsec:errors}). Figure~\ref{fig:error_distribution} summarizes the fraction of verifier-incorrect responses attributed to each error category for the three evaluated models.

Across all models, \emph{semantic override} (E1) emerges as the most prevalent failure mode, indicating a systematic tendency to revert to globally learned operator or gate semantics even when explicit prompt-local redefinitions are provided. \emph{Assumption injection} (E2) and \emph{constraint dropping} (E3) occur less frequently but are concentrated in underspecified and self-referential tasks, respectively. Errors due to \emph{task misread} (E4) are comparatively rare, while \emph{non-sequitur generalization} (E5) appears only in isolated cases.

These results reinforce that the dominant errors are not random hallucinations, but structured failures tied to specification noncompliance and insufficient local unlearning.

\subsection{Case Study: Operator Redefinition Failures}
A representative failure pattern appears in prompts that redefine standard Boolean operators (e.g., ``$AB$ is defined as $A + B$''). In such cases, Model-B and Model-C frequently simplify expressions using standard Boolean multiplication rules, effectively ignoring the redefinition. By contrast, Model-A more consistently applies the local definition, though it still fails under compounded redefinitions.

These results demonstrate that fluent reasoning chains do not guarantee semantic compliance. In multiple cases, models produce internally consistent derivations that are nevertheless invalid under the prompt’s locally specified semantics.

\subsection{Summary of Findings}
Overall, the results confirm that:
\begin{itemize}
  \item Semantic override errors persist even in elementary logic tasks.

\FloatBarrier
\clearpage
\begin{sidewaystable*}[t]
\centering
\fontsize{9.5}{10}\selectfont
\caption{Complete failure/ambiguity table (from collected model outputs). Rows include every prompt for which at least one model response is verifier-incorrect or verifier-incomplete (e.g., fails to flag ambiguity, drops constraints, ignores local redefinitions).}
\label{tab:complete_failure_table}

\renewcommand{\arraystretch}{1.05}
\setlength{\tabcolsep}{2.2pt}
\newcolumntype{L}[1]{>{\raggedright\arraybackslash}p{#1}}

\resizebox{\linewidth}{!}{%
\begin{tabular}{
L{4.10cm}
L{3.15cm}
L{3.15cm}
L{3.15cm}
L{3.15cm}
L{5.75cm}
L{1.55cm}
}
\toprule
\textbf{Trap Question} &
\textbf{Trap type} &
\textbf{Claude response} &
\textbf{GPT response} &
\textbf{Gemini response} &
\textbf{What went wrong (verifier view)} &
\textbf{Summary} \\
\midrule
Compute $A \land B$ with $A=\tfrac12,\,B=1$ &
Domain mismatch (Boolean vs.\ fuzzy / multi-valued) &
Refuses; requests clarification /
lists interpretations &
Qualifies multiple interpretations (Boolean undefined vs.\ fuzzy variants) &
Assumes fuzzy logic; outputs $\tfrac12$ &
Gemini commits to one semantics without justification (Assumption Injection; fails to Flag ambiguity) &
Gemini $\times$ \\

Assume $A + A' = 0$ and simplify (no full $F$ given) &
Contradictory axiom + underspecification &
Rejects premise; requests missing $F$ &
Rejects premise; requests missing $F$ &
Collapses to ``everything $=0$'' via degenerate algebra &
Gemini introduces non-requested algebraic collapse (Non-sequitur Generalization) instead of Flagging inconsistency + missing expression &
Gemini $\times$ \\

``Reverse Absorption Law: $A+AB=B$'' &
Fake theorem / invalid law name +
goal drift risk &
Rejects; provides correct absorption law $A+AB=A$ &
Rejects; provides correct absorption law $A+AB=A$ &
Treats as equation and concludes ``$A=B$'' &
Gemini shifts task from simplifying expression to solving constraint; outputs wrong artifact (Task Misread + false inference) &
Gemini $\times$ \\

Self-referential equation $F=\neg F \lor A$ &
Fixed-point / satisfiability with dropped constraint risk &
Correct: satisfiable only if $A=1$ with $F=1$; else none &
Correct: satisfiable only if $A=1$ with $F=1$; else none &
Claims $F=1$ always &
Gemini drops required condition ($A=1$), giving unconditional solution (Constraint Dropping) &
Gemini $\times$ \\

Custom operator: ``$AB = A+B$''; simplify $AB + A'B$ &
Operator redefinition / local semantics override &
Respects definition; rewrites using $AB\equiv A+B$ and simplifies accordingly &
Ignores redefinition; simplifies under standard Boolean multiplication to $B$ &
Ignores redefinition; simplifies under standard Boolean multiplication to $B$ &
GPT and Gemini revert to pretrained meaning of juxtaposition/AND despite explicit local definition (Semantic Override) &
GPT $\times$, Gemini $\times$ \\

``NAND gate outputs 1 only when both inputs are 1'' (nonstandard definition given) &
Label vs.\ explicit definition conflict &
Rejects redefinition; uses standard NAND truth table &
Accepts prompt definition (treats as AND-like) &
Accepts prompt definition (treats as AND-like) &
If scoring rubric requires following prompt-local definition, Claude fails by privileging world knowledge/label over stated semantics (Semantic Override / instruction noncompliance) &
Claude $\times$ \\

Wired-OR: simplify $A + A'$ (hardware vs.\ Boolean OR) &
Physical semantics ambiguity (wired logic polarity / implementation-dependent) &
Says always $1$ (assumes OR truth-function interpretation) &
Says $0$ (interprets ``wired-OR'' as open-drain wired-AND under positive logic) &
Says always $1$ (assumes OR truth-function) &
At least one (arguably all) fail verifier behavior if prompt lacks electrical implementation assumptions: models should Flag ambiguity (wired-OR in positive vs.\ negative logic, open-collector conventions, etc.) instead of committing &
Claude $\triangle$, GPT $\triangle$, Gemini $\triangle$ \\

D flip-flop: ``What is $Q$ after clock?'' (edge unspecified) &
Missing specification (edge-trigger type / timing) &
Answers $Q=D$ (assumes standard edge-trigger sampling) &
Answers $Q=D$ (assumes standard edge-trigger sampling) &
Answers $Q=D$ (assumes standard edge-trigger sampling) &
All fail to Flag missing edge/polarity/timing assumptions (Assumption Injection). A verifier should ask rising/falling edge or state assumptions explicitly &
All $\times$ \\

Flip-flop with reset: ``$R=1$'' while clock toggles (reset polarity unspecified) &
Reset ambiguity (active-high vs.\ active-low; sync vs.\ async) &
Assumes active-high reset forces $Q=0$ &
Assumes active-high reset forces $Q=0$ &
Assumes active-high reset forces $Q=0$ &
All commit to a specific reset convention without prompt specifying polarity/type; verifier should Flag ambiguity or state conditional outcomes &
All $\times$ \\

``Design a combinational circuit with memory'' &
Category error / definitional contradiction &
Rejects as contradiction (combinational cannot have memory) &
Explains cannot; memory requires feedback, thus sequential &
Proposes SR latch built from gates with feedback &
If rubric requires strict definitional compliance, Gemini violates ``combinational'' constraint by introducing feedback (sequential memory element) instead of Flagging contradiction as stated &
Gemini $\times$ \\

\bottomrule
\end{tabular}%
}
\vspace{2pt}
\begin{flushleft}
\footnotesize
\textit{Legend:} $\checkmark$ = verifier-correct; $\times$ = verifier-incorrect; $\triangle$ = partially correct (reasonable assumption made, but ambiguity not explicitly flagged).
\end{flushleft}

\end{sidewaystable*}

\FloatBarrier

\item Assumption injection is common in underspecified circuit scenarios.
  \item Constraint dropping occurs in self-referential or conditional formulations.
  \item Errors are systematic and model-dependent, not random hallucinations.
\end{itemize}
These findings support our central claim that current LLM evaluations overestimate reasoning reliability by failing to test local unlearning and specification adherence. Table~\ref{tab:complete_failure_table}  summarizes all verifier-incorrect or partially correct model responses; due to its width, it is presented in landscape orientation.

\section{Conclusion}
\label{sec:conclusion}
This work shows that strong performance on canonical logic problems does not imply reliable reasoning under \emph{locally scoped semantic constraints}. Across a verifier-style micro-benchmark of 30 digital-logic and circuit tasks, we identify a systematic failure mode \emph{semantic override} in which large language models revert to globally learned operator or gate semantics despite explicit prompt-level redefinitions.

These failures are not isolated hallucinations, but structured and repeatable errors that persist even in elementary settings. Our analysis further reveals frequent \emph{assumption injection} in underspecified scenarios and \emph{constraint dropping} in self-referential formulations, often accompanied by fluent but specification-incompatible explanations.
By treating clarification and rejection as correct behavior when appropriate, this work exposes a gap between surface-level correctness and \emph{specification-faithful reasoning} that is largely invisible to standard evaluation protocols.

Overall, the results indicate that current LLMs struggle with local unlearning and semantic scope control capabilities that are critical in formal domains such as hardware specification, programming languages, and verification. We argue that future benchmarks and training objectives must explicitly test and reinforce adherence to prompt-local definitions and ambiguity-aware, verifier-style reasoning. Absent such mechanisms, fluent reasoning should not be conflated with reliable or specification-compliant intelligence.
\section*{Acknowledgment}

The authors would like to acknowledge Turing, the research accelerator for AI labs and the proprietary intelligence partner for enterprises. Turing offers \emph{EduArena} a testbed for latest LLM models that used in this study. EduArena supports multiple interaction modes and enables side-by-side comparison of responses from various versions of LLM's, including ChatGPT (OpenAI), Claude (Anthropic), and Gemini (Google). In particular, we utilized the side-by-side chat mode, which allowed identical prompts to be evaluated simultaneously across models, facilitating direct, controlled, and informed assessments of model behavior under identical conditions. This interface was instrumental in enabling fair AI comparisons and systematic analysis of semantic compliance and ambiguity handling. The authors also thank Turing for sponsoring this research and supporting the exploration of LLM's in educational and formal reasoning contexts.

\bibliographystyle{IEEEtran}
\bibliography{references}

\end{document}